\documentclass{article}
\usepackage{amsmath,geometry,physics,amssymb,mathtools,array,tabularx,amsfonts,cite,tcolorbox,subcaption,tocloft,enumitem,textcomp,tikz,jheppub}

\definecolor{DCviolet}{RGB}{140, 43, 226}

\begin{document}

\title{Krylov Complexity and $C$-function along RG Flows}

\author[a]{Carlos Nunez}\author[b]{ and~ Dibakar Roychowdhury}
\affiliation[a]{Centre for Quantum Fields and Gravity, Department of Physics, Swansea University, Swansea SA2 8PP, United Kingdom}
\affiliation[b]{Department of Physics, Indian Institute of Technology Roorkee,
Roorkee 247667, Uttarakhand, India}
%
%
%%%%%%%%%%%%%%%%%%%%%%%%%%%%%%%
\abstract{ 
We investigate Krylov spread complexity along holographic renormalisation-group flows using the proposal that its growth rate is captured by the proper radial momentum of an infalling massive probe. We focus on the second time derivative of the complexity $\mathcal{U}$,
 which is determined locally by the redshift and radial metric functions of the dual geometry. For Lorentz-invariant domain-wall flows preserving the spacetime dimension, we derive a new relation between ${\cal U}$ and the covariant central charge $c_{\text{cov}}$.The decrease of the covariant central function towards the infrared is accompanied by a monotonic increase of the complexity acceleration. For the top-down Dp-brane family, we  obtain a universal relation. We then examine flows across dimensions, including a twisted compactification from four to two dimensions and from six to four dimensions. In these examples $c_{\rm cov}$ and $\mathcal{U }$ 
 are co-monotonic, in sharp contrast with the inverse correlation characteristic of fixed-dimensional flows. We argue that this reversal reflects the reorganisation, rather than simple depletion, of degrees of freedom into lower-dimensional sectors under compactification. Our results identify complexity acceleration as a sensitive geometric diagnostic connecting information spreading, holographic central functions and RG evolution.}
%%%%%%%%%%%%%%%%%%%%%%%%%%%%%%%%%%
%------ you can add your emails here -----------
%\emailAdd{dchatzis@proton.me}
%\emailAdd{m.hammond.2412736@swansea.ac.uk}
\emailAdd{c.nunez@swansea.ac.uk}
%\emailAdd{alfonso.ramallo@usc.es}
\emailAdd{dibakar.roychowdhury@ph.iitr.ac.in}
%\begin{document} 
\maketitle
\flushbottom
%\newpage
%%%%%%%%%%%%%%%%%%%%%%%%%%%%%%%%%%%%%%%%%%%%%%%%%%%%%%%
%\flushbottom
%\newpage
%%%%%%%%%%%%%%%%%%%%%%%%%%%%%%%%%%%%%%%%%%%%%%%%%%%%%%%

%\tableofcontents
\section{Introduction and General Idea of this paper}

Quantum complexity provides a quantitative language for describing how difficult it is to
prepare a state, implement a unitary transformation, or reproduce the dynamical growth of an
operator.  Different definitions address different questions: circuit and Nielsen
complexities measure the cost of synthesis relative to a chosen gate set and cost function,
whereas Krylov complexity is intrinsic to a Hamiltonian, an initial state or operator, and a
choice of inner product.  The latter construction is based on the Lanczos algorithm and
recasts quantum evolution as a one-dimensional hopping problem.  Comprehensive accounts of
the method, its applications and its relation to quantum chaos, many-body dynamics, quantum
field theory and holography can be found in the reviews
\cite{Nandy:2024evd,Baiguera:2025dkc,Rabinovici:2025otw}.

For definiteness, consider a normalised state
\(\ket{\psi(t)}=e^{-iHt}\ket{\psi_0}\).  Starting from
\(\ket{K_0}=\ket{\psi_0}\), the Lanczos procedure constructs an orthonormal basis in which
the Hamiltonian is tridiagonal,
\begin{equation}
 H\ket{K_n}=b_{n+1}\ket{K_{n+1}}+a_n\ket{K_n}
 +b_n\ket{K_{n-1}},\qquad b_0=0.\label{betoalonso}
\end{equation}
Expanding the evolved state as
\(\ket{\psi(t)}=\sum_{n\geq0}\varphi_n(t)\ket{K_n}\), one obtains a wavefunction on a
semi-infinite chain.  The spread complexity is the mean position of this wavefunction,
\begin{equation}
 {\cal C}_K(t)=\sum_{n=0}^{\infty}n\,|\varphi_n(t)|^2 .
\end{equation}
It measures how far the state has propagated from its initial Krylov site.  The return or
survival amplitude
\(\varphi_0(t)=\bra{\psi_0}e^{-iHt}\ket{\psi_0}\) determines the moments of the associated
spectral measure and hence, subject to the usual moment-problem assumptions, the Lanczos
data.  Thus the Krylov chain packages spectral information into a simple effective
mechanical system.  The closely related operator construction replaces \(H\) by the
Liouvillian \({\cal L}=[H,\cdot]\).  Although state spread complexity and operator Krylov
complexity share the same recursion technology, they should not be identified: they
describe propagation in different Hilbert spaces and need not contain the same dynamical
information \cite{Muck:2026top}.

The Krylov description is especially appealing in quantum field theory, where direct
control of real-time evolution is difficult.  The Lanczos coefficients determine the
effective hopping strengths along the chain, while the probability distribution
\(|\varphi_n(t)|^2\) resolves how the evolving excitation occupies successively more
complicated Krylov sectors.  Consequently, the growth, saturation, oscillation or
localisation of \({\cal C}_K(t)\) can diagnose qualitatively different dynamical regimes.
The method has been used to study chaos and integrability, confinement and
deconfinement, symmetry resolution, matrix models, conformal field theories and
gravitational systems; see again
\cite{Nandy:2024evd,Baiguera:2025dkc,Rabinovici:2025otw} and references therein.
Nevertheless, applications to renormalisation-group flows remain comparatively scarce.
In particular, it is not yet well-understood in general how the reorganisation of degrees of
freedom along an RG flow is encoded in the Lanczos coefficients or in the probability
current carried by the Krylov chain.

A decisive step towards a geometric description was made by Caputa, Chen, McDonald,
Sim\'on and Strittmatter \cite{Caputa:2024sux}.  For locally excited states in
two-dimensional CFTs with an AdS$_3$ dual, they established a precise equality, up to the
orientation convention for the radial coordinate, between the rate of spread complexity
and the proper radial momentum of the corresponding falling bulk particle.  This result
sharpens earlier momentum-complexity ideas \cite{Barbon:2020uux, Barbon:2020olv} by identifying both the appropriate notion of
complexity and the {\it proper} (rather than coordinate) momentum.  Related formulations and
extensions were developed in Refs.~\cite{Fan:2024iop,He:2024pox, Li:2025fqz}.  The resulting
prescription converts a boundary information-theoretic observable into a tractable
geodesic problem and has opened a new line of investigation in top-down holography.

This line has been developed in a sequence of works to which the present paper belongs.
The prescription was tested in a smooth confining geometry, where the finite radial
interval produces oscillatory spread complexity \cite{Fatemiabhari:2025usn}, and was then
extended to conformal quiver theories in which motion along an internal quiver coordinate
probes colour and flavour data \cite{Fatemiabhari:2025poq}.  A systematic survey of
confining backgrounds identified the confinement scale as the characteristic frequency
governing the oscillations \cite{Fatemiabhari:2026goj}.  Higher-dimensional conformal
theories, quiver motion and \(R\)-symmetry charge were studied in
\cite{Fatemiabhari:2026rob}; charged, composite and extended probes were analysed in
\cite{Nastase:2026lhz}; and genuine D-brane and string probes, together with the
fixed-charge Routhian prescription, were developed in \cite{Chatzis:2026ekd}.  These
results show that the proper-momentum proposal is sensitive not only to radial scale but
also to confinement, internal symmetries, quiver structure and the extended nature of the
boundary excitation.

The subject is developing rapidly.  Direct extensions have considered Yang--Baxter
deformations, Coulomb-branch flows, Lin--Maldacena geometries, plane-wave matrix models,
higher Krylov correlators, Lifshitz and hyperscaling-violating backgrounds, and the
universality of probe complexity
\cite{Roychowdhury:2026eds,Zoakos:2026obl,Roychowdhury:2026sgg,
Roychowdhury:2026vzq,Roychowdhury:2026igc,Alfinito:2026vah,
BitaghsirFadafan:2026lek,Graef:2026pzv,Roychowdhury:2026mpd}.
The same programme has entered analyses of observer-dependent holographic spread,
orthogonal-polynomial reconstruction, supersymmetric large-\(N\) quantum mechanics and
controlled chaos \cite{Li:2025fqz,Qu:2025lgo,Alfinito:2026yex,Baume:2026jyt}.
It has also informed work on moduli spaces, phase transitions and criticality in
supergravity backgrounds \cite{Anabalon:2025sok,Anabalon:2026yxk,
Elander:2026eyk,Anabalon:2026zxp}.  Recent conceptual discussions have clarified both the
information encoded by Krylov observables and the assumptions required by the bulk
prescription \cite{Muck:2026top,Li:2026pdh}.  This growing body of work makes it timely to
ask whether the radial dependence of spread complexity can be related directly to a
standard measure of the number of field-theory degrees of freedom.

\subsection{Organisation and main results}

The purpose of this paper is to study spread Krylov complexity along holographic RG flows
and to compare its acceleration with a suitably defined covariant holographic \(c\)-function.  We define
\begin{equation}
 {\cal U}(t)=\frac{2}{m\sqrt{A(r_{\rm UV})}}\,\ddot{\cal C}_K(t),
\end{equation}
where \(A(r)\) is the redshift factor seen by a massive probe released from rest at
\(r_{\rm UV}\).  Our central result is that \({\cal U}\) and the covariant central function
are not unrelated diagnostics: in every class of flows considered here they obey a direct
algebraic or parametric relation.  The character of that relation, however, depends
sharply on whether the flow preserves the number of spacetime dimensions or is generated
by compactification.  To the best of our knowledge, this is the first systematic
comparison between a holographic \(c\)-function and the second time derivative of spread
complexity along RG flows.

In Section~\ref{expresionesgenerales} we collect the general formulas used throughout the paper.  For a massive
particle following a consistently truncated radial geodesic, we derive the trajectory, the
proper momentum, \({\cal C}_K\), \(\dot{\cal C}_K\) and \({\cal U}\) in terms of the metric
functions \(A(r)\) and \(B(r)\).  We then review the covariant \(c\)-function of
Ref.~\cite{Jokela:2026cjm}, which can be applied directly in ten or eleven dimensions and
therefore retains information about internal manifolds and fibrations.

In Section~\ref{RGflowssame} we study Lorentz-invariant domain-wall flows that preserve the spacetime
dimension.  For metrics in $(d+1)$ dimension of the form \(ds^2=e^{2a(r)}dx_{1,d-1}^2+dr^2\), we find the new and remarkably
simple relation (here $G_N^{(d+1)}$ is the Newton constant)
\begin{equation}
 c_{\rm cov}(r)=\frac{1}{G_N^{(d+1)}\,{\cal U}(r)^{d-1}} .
\end{equation}
The same energy condition that makes \(c_{\rm cov}\) decrease towards the infrared makes
\({\cal U}\) increase along a falling trajectory.  Hence the loss of effective degrees of
freedom is accompanied by an increasing acceleration of the Krylov spread.  We illustrate
the statement with the GPPZ flow and discuss the limitations associated with its singular
infrared endpoint.

Section~\ref{sectiondpbranesflow} provides a top-down test using the near-horizon geometries of D\(p\)-branes.  We
solve the radial problem, analyse the special values of \(p\), and obtain
\begin{equation}
c_{\rm cov}~{\cal U}^{8}=\Upsilon(p),\nonumber
\end{equation}
where \(\Upsilon(p)\) is independent of the radial
coordinate.  The exponent eight is common to the entire D\(p\)-brane family, while the
normalisation remembers \(p\), the Yang--Mills coupling and the number of colours.  This
generalises the inverse relation found for domain walls and makes explicit that the
central function and the complexity acceleration probe complementary aspects of the same
radial evolution.

In Section~\ref{sectionRGflowacross} we turn to flows across dimensions.  We analyse a twisted compactification
from four to two dimensions and the wrapped-M5 flow from six to four dimensions.  In both
examples the covariant \(c\)-function and \({\cal U}\) are co-monotonic, in marked contrast
with the inverse relation characterising same-dimensional domain walls.  We derive an
explicit algebraic relation in the first example and a parametric relation, together with
controlled ultraviolet and infrared expansions, in the second.  This reversal is another of
the  new observations of this work.  It reflects the fact that compactification
changes the dimensionality and Lorentz group of the effective theory and reorganises
lower-dimensional massless sectors; consequently, the covariant central function is not
constrained to behave as the \(c\)-function of a Lorentz-invariant flow within a fixed
dimension.  The comparison therefore suggests that the acceleration of spread complexity
is a sensitive diagnostic of how degrees of freedom are reorganised, rather than merely
removed, along an RG trajectory.
Appendix \ref{appendixa} hints at a connection between the calculations in supergravity and the Lanczos coefficients. This connection should be further explored in the future.

\section{Holographic expressions for spread complexity and c-function}\label{expresionesgenerales}
In this section we summarise some expressions that are used in the rest of the paper. These are taken from the papers \cite{Fatemiabhari:2026goj, nunez:2026lio}.
We consider generic backgrounds of the form
\begin{equation}
ds^2= \widehat{A}(r,\vec{y})\left(-dt^2+ d\vec{x}_{d-1}^2 + \widehat{B}(r,\vec{y})dr^2\right) + g_{ij}(r,\vec{y})dy^i dy^j.\label{genericbackground}    
\end{equation}
These are (generically) holographic duals to relativistic QFTs in $d$-spacetime dimensions (in this work we usually consider backgrounds in ten or eleven dimensions). The $r$-dependence indicates an RG-flow and the $y^i$-dependence is usually associated with internal symmetries of the QFT, or a quiver-like structure or more elaborated dynamics. Below, we discuss examples. 

We summarise expressions needed to compute the holographic dual of the spread Krylov complexity. To do this, we  follow \cite{Caputa:2024sux, Fan:2024iop, He:2024pox} and consider  a massive point-like particle that falls radially in a generic background like (\ref{genericbackground}). In the treatment we present below, it is {\it assumed} that it is possible to fix the coordinates $(\vec{x}=\vec{x}_0,\vec{y}=\vec{y}_0)$ to constant values. If this is not the case, the dynamical equations become more complicated and we are in situations like the one studied for quiver gauge theories in  \cite{Fatemiabhari:2025poq, Fatemiabhari:2026rob}, for which the simplified expressions below, do not apply. 

Under these conditions, the motion of a radially in-falling object can be parametrised by $r(t)$. We fix all other coordinates to constant (in a way consistent with the equations of motion). The corresponding induced metric on the world-line of the particle is 
\begin{equation}
 d s_{\text{{ind}}}^2= \left(-A(r) + A(r) B(r)\dot{r}^2\right)d t^2, ~~A(r)=\widehat{A}(r,\vec{y}_0),~~B(r)=\widehat{B}(r,\vec{y}_0).\label{specialxx}
\end{equation}
The dynamics is thus governed by the world-line action
\begin{equation}
 S =-m \int \dd t \sqrt{A(r)\left( 1-B(r)\dot{r}^2\right)}
\, \equiv \, \int \dd t \, L\,.
  \label{action-KS}   
\end{equation}
The last equality defines the Lagrangian $L$, the dot denotes derivative with respect to time, and $m$ is the mass of the particle. The concrete expressions of $A(r)$ and $B(r)$  are example-dependent. Different examples are studied in the coming sections. 

The equation of motion for $r(t)$ reads,
\begin{equation}
 \frac{\dd}{\dd t}\Bigg[ \frac{A B \dot{r}}{L}\Bigg]= \frac{-A'(1- B\dot{r}^2) +A B'\dot{r}^2}{2L},
 \label{eq-mot-r}  
\end{equation}
with primes indicating derivatives with respect to $r$. Since the integrand in Eq.~\eqref{action-KS} does not depend on time, the Hamiltonian
\begin{equation}
H= \dot r \frac{\partial L}{\partial \dot r} - L= m \frac{A}{\sqrt{A(1- B \dot{r}^2)}} \label{lionel} 
\end{equation}
is conserved. Following \cite{Caputa:2024sux}, we choose as initial conditions that the particle is  released at $t=0$ from $r(0)=r_{UV}$, where it is initially at rest, that is $\dot r(0) = 0$. With this choice, the Hamiltonian~\eqref{lionel} is simply
\begin{equation}\label{eq:Hmailtonian_particle}
H= m\sqrt{A(r_{UV})}\,,
\end{equation}
and  a first integral of the equation of motion is found from eq.~\eqref{lionel},
\begin{equation}\label{eq:rdot_Krylov}
 \dot{r}=\pm \sqrt{\frac{1}{B(r)}\left[1-\frac{A(r)}{A(r_{UV})} \right]}\,.
\end{equation}
%
%To develop some intuition, it is useful to recast the problem in terms of the motion of the particle in an effective potential, rather than as geodesic motion in a curved spacetime. To this end, we write the Hamiltonian in Eq.~\eqref{lionel} in terms of the momentum conjugate to $r$, namely
%\begin{equation}
 %   P_r=\frac{\partial L}{\partial\dot{r}}= m \sqrt{\frac{A(r) B(r)^2~\dot{r}^2}{1- B(r) ~\dot{r}^2}}\,.
%\end{equation}
%Then
%\begin{equation}
%H^2= m^2 A(r) + \frac{P_r^2}{B(r)}. \label{PrH2}
%\end{equation}
%Consequently, in {\it analogy} with the Newtonian energy balance, the term $m^2A(r)$ plays the role of an effective potential, while $B(r)$ controls the radial effective mass of the particle.
%
%
In Refs.~\cite{Caputa:2024sux, He:2024pox, Fan:2024iop}, the time derivative of the Krylov complexity ${\cal C}_K(t)$ was identified with the proper momentum of the particle, which we denote below by $P_{\overline{y}}$.  
The proper-coordinate $\bar{y}$ is defined through the metric 
\begin{equation}
\dd s^2= A(r) B(r) \dd r^2= \dd {\bar{y}}^2 \quad \Rightarrow\quad  
\frac{\dd\dot{r}}{\dd\dot{\bar{y}}}=\pm  \frac{1}{\sqrt{AB}}.
\end{equation}
In terms of this variable, the Lagrangian in Eq.~\eqref{action-KS} reads
\begin{equation}
    L=-m \sqrt{A(r)-\dot{\bar{y}}^2}\,.
\end{equation}
Then
\begin{eqnarray}
& & \dot{\cal C}_K(t) =- P_{\bar{y}}=-\frac{\partial L}{\partial \dot{r}} \frac{d\dot{r}}{d\dot{\bar{y}}}=- 
\frac{m A B  ~\dot{r}}{\sqrt{A (1- B \dot{r}^2)} }\times\frac{1}{\sqrt{A B}}=- m \dot{r} \sqrt{\frac{B}{1- B\dot{r}^2}}~.\label{complexityBBgeneric}
\end{eqnarray}
Notice that time derivative of the complexity can be positive or negative. One finds that when the particle is {\it falling} and $\dot{r}<0$, the complexity grows ($\dot{\cal C}_K(t)>0$) and when the particle climbs-up ($\dot{r}>0$) the complexity decreases ($\dot{\cal C}_K(t)<0$). In this work we consider situations in which the probe particle never climbs-up the background. Those situations are found in \cite{Fatemiabhari:2025usn, Fatemiabhari:2026goj, Zoakos:2026obl, nunez:2026lio}.

The Krylov spread complexity ${\cal C}_K(t)$ has no units. 
Putting the expressions above (\ref{eq:rdot_Krylov}) together (and using that the particle starts falling from $r_{UV}$ with zero initial velocity), we find
\begin{equation}
 \dot{\cal C}_K(t)=  m~  \sqrt{\frac{A(r_{UV})}{A(r)} -1}.   \label{cdot}
\end{equation}
To account for the cases in which the particle climbs-up the background--see the discussion below eq.(\ref{complexityBBgeneric}), the paper \cite{nunez:2026lio} adds a $-\text{sign}(\dot{r})$ to eq.(\ref{cdot}). For this simple expression to be useful, $r(t)$ must be found first, solving eq.(\ref{eq:rdot_Krylov}). 
We take a time-derivative of eq.(\ref{cdot}), apply the  chain rule and use eq.(\ref{eq:rdot_Krylov}) and find,
\begin{equation}
 \ddot{\cal C}_K=\frac{m}{2}\frac{\sqrt{A(r_{UV})}A'(r)}{\sqrt{A^3(r)B(r)}}.    \label{cdotdot}
\end{equation}
It is convenient for what follows to define the quantity,
\begin{equation}
{\cal U}=\frac{2}{m\sqrt{A(r_{UV})}} \ddot{\cal C}_K(t)=\frac{A'(r)}{\sqrt{A^3(r) B(r)}}.    \label{udet}
\end{equation}
Finally, we can derive an implicit expression for the spread complexity. Writing eq.(\ref{cdot}) as
\begin{eqnarray}
& & \frac{1}{m} \frac{d ~{\cal C}_K}{dr} \times \frac{dr}{dt}= \sqrt{\frac{A(r_{UV})}{A(r)}-1},~~\text{using eq.(\ref{eq:rdot_Krylov}) we find,}\nonumber\\
& & \frac{1}{m \sqrt{A(r_{UV})}} {\cal C}_K(t)=\int_{r(t)}^{r_{UV}} dr \sqrt{\frac{B(r)}{A(r)}}.\label{complexityrt}
\end{eqnarray}
Notice that ${\cal C}_K(0)=0$ as $r(0)=r_{UV}$. As mentioned above, these expressions are useful if we first solve for $r(t)$ using eq.(\ref{eq:rdot_Krylov}) and then evaluate (\ref{cdot}),(\ref{udet}),(\ref{complexityrt}).
\subsection*{\underline{\bf An interesting special case}}
Let us consider the particular situation for which $A(r) B(r)=1$. This is a gauge choice $\sqrt{A(r) B(r)} ~dr=d\rho$, but finding $r(\rho)$ might be difficult. Once we know $r(t)$, or $\rho(t)$ by solving eq.(\ref{eq:rdot_Krylov}) we can write,
\begin{equation}
{\cal C}_K(t)=m \sqrt{A(r_{UV})} \int_{r(t)}^{r_{UV}} \frac{dr}{A(r)}, ~~  \dot{\cal C}_K(t)= m\sqrt{\frac{A(r_{UV})}{A(r)} -1},~~~{\cal U}(t)=\frac{A'}{A}.\label{formulasutiles}    
\end{equation}
In forthcoming sections, we evaluate these quantities for different holographic backgrounds.
Let us now summarise a second ingredient of our study, {\it the holographic covariant c-function}.
\subsection{Holographic covariant c-function}
There is a long history of attempts to construct quantities in supergravity mirroring the behaviour  
of a c-function in QFT. The c-function, like the beta-function is only well defined at conformal fixed points (otherwise, it is scheme dependent). The idea from the perspective of supergravity is to write a quantity that at the fixed points takes a constant value associated with the central charge (free energy) of the dual CFT and along the flow is monotonic, decreasing towards the IR for Lorentz invariant situations. See the papers \cite{Freedman:1999gp, Alvarez:1998wr,Sahakian:1999bd, Macpherson:2014eza, Bea:2015fja, Merrikin:2022yho, Jokela:2026cjm, Caceres:2023mqz} as representative of various attempts in this direction. We adopt the definitions and notation of \cite{Jokela:2026cjm}. Consider a background for a theory of gravity in $(d+1+p)$ dimensions that is holographic dual to a QFT in $d$-dimensions. The Einstein-frame metric is a generalisation of that in  eq.(\ref{genericbackground}),
\begin{eqnarray}
& & ds^2_{d+1+p} = -\mathcal{N}^2 dt^2  +g_{ab}\,dx^adx^b+ g_{rr}\,dr^2 + h_{ij}\,(dy^i + A^i_a\,dx^a)(dy^j + A^j_b\,dx^b),\label{genericbackgroundcentral}\\
& & 
\mathcal{N} = \mathcal{N}(r,\vec{y})\,,\quad  g_{ab} = g_{ab}(r,\vec{y})\,,\quad g_{rr} = g_{rr}(r,\vec{y})\,,\quad h_{ij} = h_{ij}(r,\vec{y})\,,\quad A^i_a = A^i_a(r,\vec{y}).\nonumber\\
& & a,b=1,....,d-1;~~~~i,j=1,....,p.\nonumber
\end{eqnarray}
The c-function defined in \cite{Jokela:2026cjm} is written in terms of the extrinsic curvature of a space-like slice of the background. It is a covariant definition. After various manipulations carefully described in \cite{Jokela:2026cjm} one can write an expression in terms of the metric (\ref{genericbackgroundcentral}). It reads
\begin{eqnarray}
& & {c}_{\text{cov}}(r)=\frac{(d-1)^{d-1}}{G_{\mathrm N}^{(d+1+p)}}
\int d^p y\,\sqrt{h}\,
\biggl(\frac{\sqrt{g_{rr}}}{\partial_r \log(gh)}\biggr)^{d-1},\label{cfunction}
\\
& & h = \det h_{ij}\,,\quad g = \det g_{ab}.\nonumber
\end{eqnarray}
In the forthcoming sections, we calculate the quantities in eqs.(\ref{cdot})-(\ref{formulasutiles}) and (\ref{cfunction}) for different backgrounds, and establish relations among them.
\section{RG-flows in the same space-time dimension}\label{RGflowssame}
Let us consider QFTs with a holographic dual, that undergo a renormalisation-group flow. The holographic background in eq.(\ref{genericbackground}) represents such RG flow in the QFT. This can occasionally have  AdS-fixed point(s), though this is not a necessity. Consider first a {\it bottom-up} construction of such systems, described by a solution to the equations of motion of a (gauged) supergravity in $D$-dimensions. The Lagrangian is written in terms of a metric and a set of scalar fields $\vec{\Phi}$ and reads,
\begin{equation}
L_D=\sqrt{-g}\Big(R-\frac12 G_{IJ}(\vec{\Phi})\partial\Phi^I\partial\Phi^J -V(\vec{\Phi})\Big).\label{lagrangiangauged}   
\end{equation}
There could be gauge fields and higher form fields in such lagrangian (they are ubiquitous for gauged supergravities in diverse dimensions). We consider the truncation in (\ref{lagrangiangauged}), as otherwise Lorentz invariance of the $D-1$ dimensional dual-QFT would be broken. We propose a background solution of the form,
\begin{eqnarray}
& & ds_D^2= e^{2a(r)}\left(-dt^2+ d\vec{x}^2_{D-2} \right)+ dr^2.~~~{\Phi^I(r)}.\label{metricaD}
\end{eqnarray}
Comparing with expressions (\ref{genericbackground}), (\ref{genericbackgroundcentral}), we set $p=0$ (no internal $\vec{y}$-directions) and 
\begin{equation}
{\cal N}^2= g_{ab}=\widehat{A}=\widehat{B}^{-1}= e^{2 a(r)},~~~g_{rr}=1.\nonumber
\end{equation}
Using eq.(\ref{formulasutiles}), the expressions for the Krylov spread complexity and its time derivatives are,
\begin{eqnarray}
& & \frac{{\cal C}_K(t)}{m~ e^{a(r_{UV})}}\!=\! \int_{r(t)}^{r_{UV}} dr~ e^{-2 a(r)},~~\frac{{\dot{\cal C}}_K(t)}{m}\!=\! \sqrt{e^{2\left[a(r_{UV}) - a(r)\right]} -1}~,~~\frac{2\ddot{\cal C}_K(t)}{m ~e^{a(r_{UV})}}={\cal U}(t)\!=\! 2 a'(r) .\label{expressionsflowbu}
\end{eqnarray}
These expressions become useful once we explicitly solve for $a(r)$, using a particular form for $V(\vec{\Phi})$ and $G_{IJ}(\vec{\Phi})$ together with the Einstein and dilaton equations derived from (\ref{lagrangiangauged}). Then, we  need the explicit form of the geodesic $r(t)$, found  from eq.(\ref{eq:rdot_Krylov}) to be,
\begin{equation}
 t-t_0=\int_{r}^{r_{UV}} \frac{dr}{\sqrt{ e^{2a(r)} \Big( 1- e^{2 a(r) - 2 a(r_{UV})}\Big)}}. \label{leandro}  
\end{equation}
The covariant c-function is easily obtained using eq.(\ref{cfunction}) together with $g=\det g_{ab}= e^{2(d-1)a(r)}$, the fact that $p=0$ and $D=d+1$,
\begin{equation}
c_{\text{cov}} = \frac{1}{G_{N}^{(d+1)}}\frac{1}{\Big( 2 a'\Big)^{(d-1)}} =   \frac{1}{G_{N}^{(d+1)}}\frac{1}{ {\cal U}^{(d-1)}}.\label{firstinstance}
\end{equation}
In these bottom-up set-ups the covariant c-function and the 'acceleration of the complexity' ${\cal U}(t)$ satisfy an inverse relation (\ref{firstinstance}). This relation turns out to be non-universal, as we find studying different system in what follows. The way to read eq.(\ref{firstinstance}) is the following: the closer we move to the IR-region of the QFT (smaller values of $r$), the less is the c-function and bigger is the second derivative of the complexity. In other words, the more we 'fall' the more the acceleration of the complexity grows \cite{Susskind:2018tei,Susskind:2019ddc, Ageev:2018msv}. In gapped systems like those studied in \cite{Fatemiabhari:2025usn, Fatemiabhari:2026goj, nunez:2026lio}, the second derivative of the complexity diverges (for particles with zero angular momentum), and consequently, the c-function vanishes, reflecting the gapped character of the QFT. 

Let us make the comments above a bit more precise and formal. In \cite{Freedman:1999gp}, the authors showed that for Lagrangians of the form (\ref{lagrangiangauged}) and domain wall like solutions as in eq.(\ref{metricaD}) it is possible to prove that $a''(r) <0$ if the weak energy condition is imposed. In fact, they show that
\begin{equation}
a''(r)=-\frac{1}{2(d-1)}G_{IJ}~\partial\Phi^I\partial \Phi^J.\label{weakenergy}    
\end{equation}
For the dimension of the gauged supergravity $D=d+1>2$ and using $G_{IJ}\partial\Phi^I\partial \Phi^J>0$, we have $a''<0$. This implies the monotonicity of $c_{\text{cov}}$ in eq.(\ref{firstinstance}). Indeed,
\begin{equation}
\frac{d c_{\text{cov}}}{dr}=\frac{(1-d)}{2^{d-1} G_N^{(d+1)}} \frac{a''}{\Big(a'(r)\Big)^{d}}>0.  
\end{equation}
This is used in \cite{Freedman:1999gp} to deduce that the c-function in eq.(\ref{firstinstance}) is monotonic and decreasing towards the IR.

Now, we take the independently defined quantity ${\cal U}(t)$--see eq.(\ref{udet}) and (\ref{formulasutiles}) for these domain wall geometries. We take the time derivative, that is the third derivative of the complexity and use the (negative branch of) eq.(\ref{eq:rdot_Krylov}), as the particle is falling. We find
\begin{equation}
\frac{d{\cal U}}{dt}=\frac{d{\cal U}}{dr} \times \frac{dr}{dt}= 2  a''(r) \times \Big(- \sqrt{e^{2a(r)}\left(1- e^{2a(r)-2a(r_{UV})} \right)} \Big) >0.
\end{equation}

In other words, the condition in eq.(\ref{weakenergy}) that makes the c-function monotonic and decreasing as we move towards the IR, makes the acceleration of the complexity monotonic in time (increasing as time passes). This also indicates that as long as the particle falls, the quantity ${\cal U}(t)$ increases.
For situations in which the probe particle reaches the end of the space and bounces back, see, for example, \cite{Fatemiabhari:2025usn, Fatemiabhari:2026goj, Zoakos:2026obl}, the quantity $\cal U$ will decrease as  the particle climbs up the background.
\subsubsection*{\underline{Example: Girardello Petrini Porrati Zaffaroni (GPPZ) flow}}
We consider $\mathcal{N} = 1$ SUSY preserving deformations of $\mathcal{N} = 4$ SU$(N)$ super-Yang-Mills theory in four dimensions.  This SUSY field theory is written in terms of a vector multiplet and three chiral multiplets. We denote these chiral multiplets as $(\Phi_1,\Phi_2,\Phi_3)$, not to be confused with the bulk dilaton field with a similar notation. The deformation of $\mathcal{N}=4$ SYM is written in terms of these chiral multiplets by an addition to the superpotential of the form,
	\begin{equation}
		\Delta\mathcal{W} = m_1\textbf{tr} (\Phi_1^2) +m_2\textbf{tr} (\Phi_2^2)+m_3\textbf{tr} (\Phi_3^2)\,.
	\end{equation}
	This deformation gives masses to the three chiral multiplets. When $m_1=m_2\neq 0$ but $m_3 = 0$ the supersymmetry is enhanced to $\mathcal{N} = 2$ and the corresponding flow is the Pilch--Warner flow \cite{Pilch:2000ue}. When only one chiral multiplet is non-zero the theory flows in the IR to the Leigh--Strassler conformal fixed point \cite{Leigh:1995ep}. When all three masses are non-zero, the theory is called $\mathcal{N} = 1^*$ and has a rich, well-studied structure of vacua. GPPZ is an early attempt to describe the $\mathcal{N} = 1^*$ theory using five-dimensional supergravity \cite{Girardello:1999bd}. It has not been until recently that the ten-dimensional uplift of GPPZ has been understood, see the relevant references~\cite{Petrini:2018pjk, Bobev:2018eer,Pilch:2000fu,Baguet:2015sma}.
	
	The background considered by GPPZ can be obtained from the truncated action
	\begin{equation}\label{eq:actionGH2}
		I\,=\,\frac{1}{4\pi G_5}\int d^5x\sqrt{-g}\left(\frac{R}{4}-\frac12 \partial_M\phi\partial^M\phi
		-V(\phi)\right)\,, 
	\end{equation} 
    where $M=0,\ldots,4$, the scalar potential $V(\phi)$ is given in terms of a superpotential $W(\phi)$, 
	\begin{equation}
		\label{eq:pot.from.superpot}
		V(\phi)= -\frac{4}{3}W(\phi)^2 + \frac{1}{2}\left(\frac{\partial W(\phi)}{\partial \phi}\right)^2 \ ,~~W(\phi) = -\frac{3}{4L}\left(1+\cosh{\left(\frac{2\phi}{\sqrt{3}}\right)}\right)\,.
	\end{equation}
Here, $L$ is the AdS-radius. For a domain wall background,
	\begin{equation}
		d s_5^2 = e^{2a(\rho)}(-d t^2 + d \vec{x}_3^2) + d \rho^2 \ ,~~~\phi(\rho),
	\end{equation}
	the ground state of the system is given by the solution to the BPS equations,
	\begin{equation}\label{eq:BPS}
		\partial_\rho \phi = \partial_\phi W\,,\quad \partial_\rho a = - \frac{2}{3} W\,.
	\end{equation}
In terms of the integration constant $\Lambda$, the solution reads
	\begin{equation}
		\phi(\rho) = \sqrt{3}\,\text{arctanh}\left(\frac{e^{-\rho}\Lambda}{\sqrt{3}}\right)\,, \qquad
		e^{2 a} = e^{2\rho} -{\frac{\Lambda^ 2}{3}}\,.
	\end{equation}
For $\rho\to +\infty$, $e^{2a}\sim e^{2\rho}$ and $\phi\to 0$. For $e^{2\rho_*}=\frac{\Lambda^2}{3}$, the space ends, in a singular fashion (with divergent invariants, as the divergent dilaton indicates). In other words, we can {\it only} trust our calculation far from the point $\rho_*$.

\begin{figure}
    \centering
    \includegraphics[width=0.9\linewidth]{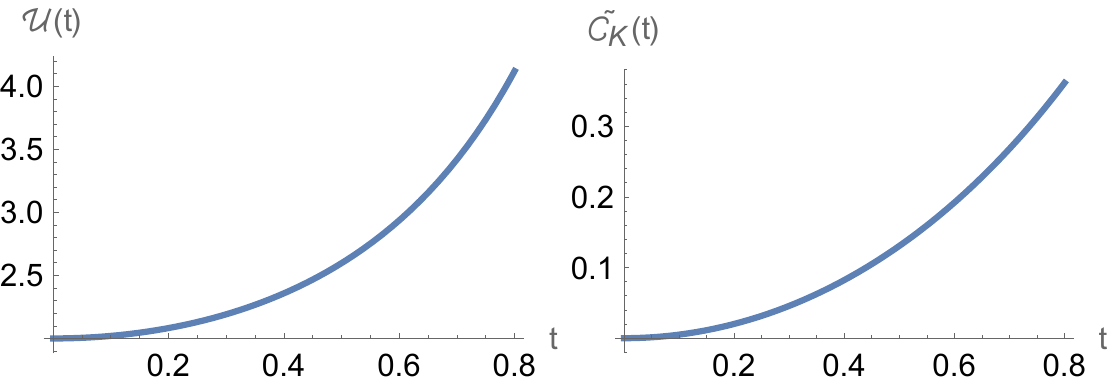}
    \caption{We plot $\mathcal{U}(t)$ and rescaled complexity $\tilde{\mathcal{C}}_K(t)$ against $t$. We set $\Lambda=\sqrt{3}$ and $\rho_{UV}=10$.}
    \label{figgppz}
\end{figure}

\begin{figure}
    \centering
    \includegraphics[width=0.8\linewidth]{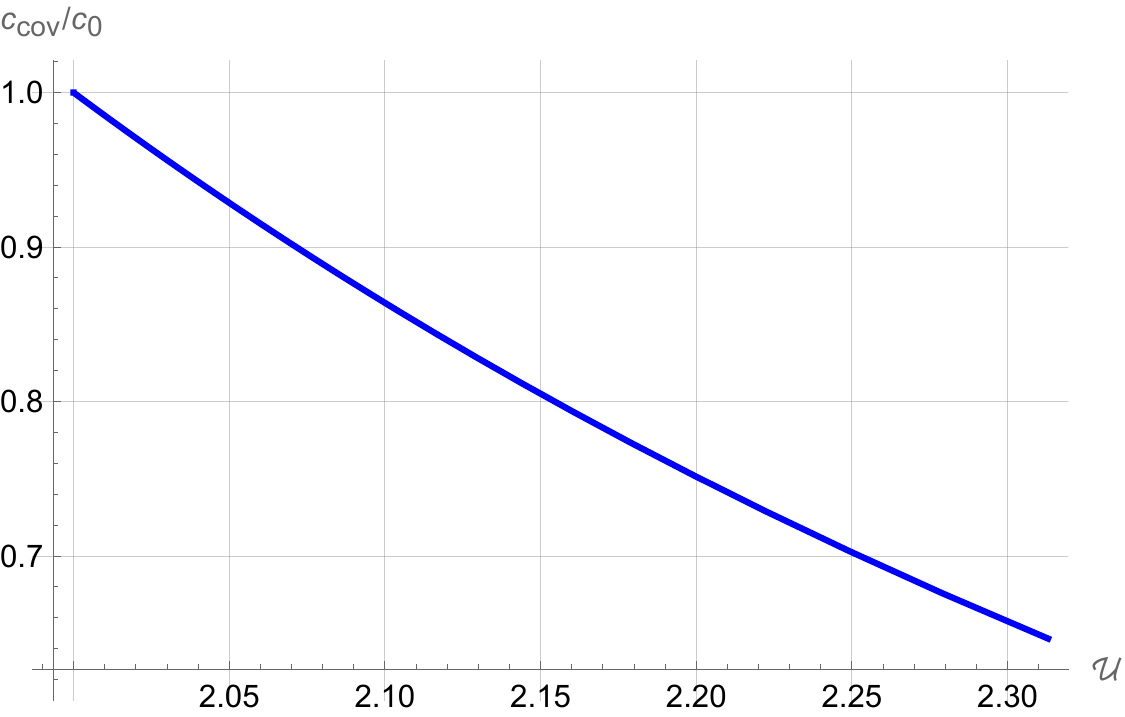}
    \caption{We plot $c_{cov}/c_0$ against $\mathcal{U}$. We set $\Lambda=\sqrt{3}$ and $\rho_{UV}=10$.}
    \label{figgppzzccov}
\end{figure}

We use eqs.(\ref{expressionsflowbu}),(\ref{leandro}),(\ref{firstinstance}) to compute the quantities,
\begin{eqnarray}
  & & \frac{{\cal C}_K(t)}{m e^{a(\rho_{UV})}}= \int_{\rho(t)}^{\rho_{UV}} dz e^{-2 a(z)}= \frac{3}{2\Lambda^2}\log\Big[ \frac{(1-\frac{\Lambda^2}{3} e^{-2\rho_{UV}})}{(1-\frac{\Lambda^2}{3} e^{-2\rho(t))}}\Big] \\
  & & \frac{\dot{\cal C}_K(t)}{m}=\sqrt{\frac{e^{2\rho_{UV}}-\frac{\Lambda^2}{3}}{e^{2\rho(t)}-\frac{\Lambda^2}{3}}-1 },\\
  & & \frac{2\ddot{\cal C}_K(t)}{m e^{a(\rho_{UV})}}={\cal U}(t)=2\frac{e^{2\rho(t)}}{\left(e^{2\rho(t)}-\frac{\Lambda^2}{3}\right)}, \\
  & & c_{\text{cov}}= \frac{1}{8 G_N^{(5)}}\left(1- \frac{\Lambda^2}{3} e^{-2\rho} \right)^3.
\end{eqnarray}
See Fig.\ref{figgppz} and Fig.\ref{figgppzzccov} for their respective variations.

The function $\rho(t)$ is found by inverting the relation,
\begin{equation}
\frac{t-t_0}{\sqrt{e^{2 \rho_{UV}} -\frac{\Lambda^2}{3}}}=\int_{\rho}^{\rho_{UV}} \frac{dz}{\sqrt{\left( e^{2z}-\frac{\Lambda^2}{3}\right)\left( e^{2\rho_{UV} } - e^{2z}\right)}}
\end{equation}
It could be of interest to study the flow by Gubser \cite{Gubser:1999pk} and Kehagias-Sfetsos \cite{Kehagias:1999tr} or the domain walls in \cite{Freedman:2003ax}. These could show subtle differences respect to GPPZ. The complexity and its derivatives are  probes to distinguish these flows from others.

Let us now study other RG flows that preserve the number of spacetime dimensions, but are phrased in explicitly top-down language, as these are formulated in a Type II (A and B) supergravity or in M-theory.
%
%\textcolor{blue}{DR: shall we make a separate section on gauge super-gravity above? as we are discussing Dp branes after this. A precise explanation of (3.5) is needed.}
%
\section{Flows in the same spacetime dimension II: the case of Dp-branes}\label{sectiondpbranesflow}
Here we study our expressions (\ref{cdot}), (\ref{udet}), (\ref{complexityrt}) and (\ref{cfunction}) in the case of Dp branes. When written in Einstein frame (appropriate to study the motion of the in-falling particle and the covariant c-function), the metric and dilaton for a Dp-brane read,
\begin{align}
    &ds^2_{\text{Einstein}}=e^{-\frac{\Phi}{2}}\Big[\hat{h}^{-1/2}(r)dx_{1,p}^2+\hat{h}^{1/2}(r) (dr^2+r^2 d\Omega_{8-p})\Big]\label{bckdp}\\
    & \hat{h}(r)=\Big( \frac{l}{r}\Big)^{7-p}~;~e^{-\frac{\Phi}{2}}=\hat{h}^{\frac{(p-3)}{8}}\nonumber.
\end{align}
We have used the convention $g_s=\alpha'=1$. The warp factors do not depend on the internal coordinates (the angles of the ($8-p$)-sphere) and the expressions in eqs.(\ref{specialxx})-(\ref{udet}) apply straightforwardly.
The required functions are
\begin{equation}
    A(r)=e^{-\frac{\Phi}{2}}\hat{h}^{-1/2}~;~B(r)=\hat{h}.  
\end{equation}
We use eqs.(\ref{cdot}),(\ref{cdotdot}),(\ref{complexityrt}) to write,
\begin{eqnarray}
& &   \frac{1}{m ~r_{UV}~\sqrt{A(r_{UV})}}\left(\frac{r_{UV}}{l}\right)^{\frac{(7-p)(15-p)}{16}} ~{\cal C}_K(t)= \int _{r(t)}^{r_{UV}} \frac{dr}{r_{UV}} \left(\frac{r_{UV}}{r}\right)^{\frac{(7-p)(15-p)}{16}},\label{complexitydpint}\\
&  &\frac{\dot{\cal C}_K}{m}=\sqrt{\Big(\frac{r_{UV}}{r} \Big)^{\frac{(7-p)^2}{8}}-1}\label{cdotdp}\\
&    &\frac{2\ddot{\cal C}_K}{m \sqrt{A(r_{UV}})}={\cal U}=\frac{(7-p)^2}{8 l}\Big( \frac{l}{r_{UV}}\Big)^{\frac{(p-3)^2}{16}}\Big( \frac{r_{UV}}{r}\Big)^{\frac{(p-3)^2}{16}}.\label{cddotdp}
\end{eqnarray}
Using eq.(\ref{eq:rdot_Krylov}), we find
\begin{equation}
    \dot{r}=-\sqrt{\left(\frac{r}{l}\right)^{(7-p)}\Bigg(1-  \left(\frac{r}{r_{UV}}\right)^{\frac{(7-p)^2}{8}}\Bigg)}.
\end{equation}
We define dimensionless quantities $x=\frac{r}{r_{UV}}$ and $\tau=\left(\frac{l}{r_{UV}}\right)^{\frac{(p-7)}{2}} \frac{t}{r_{UV}}$. In terms of these, the equation of motion  and its solution read
\begin{eqnarray}
& & \frac{dx}{d\tau}=-\sqrt{x^{(7-p)} \Big[ 1- x^{\frac{(7-p)^2}{8}}\Big]}\longrightarrow\tau-\tau_0=- \int \frac{dx}{\sqrt{x^{(7-p)} \Big[ 1- x^{\frac{(7-p)^2}{8}}\Big]}}\label{solutiondpbranes}\\
& & \tau-\tau_0=\nonumber\\
& &\frac{2}{(5-p)}\Bigg( x^{\frac{(p-5)}{2}}~{}_2F_1\big[\frac{1}{2}, \frac{4(p-5)}{(7-p)^2}, \frac{29-10p+p^2}{(7-p)^2}, x^{\frac{(7-p)^2}{8}} \big] - {}_2F_1\big[\frac{1}{2}, \frac{4(p-5)}{(7-p)^2}, \frac{29-10p+p^2}{(7-p)^2},1 \big] \Bigg).\nonumber
\end{eqnarray}
%\begin{align}
 %  \int_1^x \frac{dx}{\sqrt{(1-x^{\frac{(7-p)^2}{8}})x^{(7-p)}}} =-\frac{t}{r_{UV}}\Big( %\frac{r_{UV}}{l}\Big)^{\frac{(7-p)}{2}}+c_0.
%\end{align}
%
%\textcolor{red}{CN: please check me the previous integrals, the result you had was
%\begin{align}
%\label{e4.10}
 %   &\frac{2}{(5-p)}x^{(p-5)/2}\sqrt{1-x^{\frac{1}{8} (7-p)^2}}\nonumber\\
  %  &\times \, _2F_1\left(1,\frac{(p-3)^2}{2 (p-7)^2};\frac{(p-10) p+29}{(p-7)^2};x^{\frac{1}{8} (p-7)^2}\right)=\frac{t}{r_{UV}}\Big( \frac{r_{UV}}{l}\Big)^{\frac{(7-p)}{2}}-c_0
%\end{align}
%}

To satisfy the boundary condition $x(\tau=0)=1$, conversely $r(t=0)=r_{UV}$, we choose the integration constant
\begin{equation*}
\tau_0= \frac{2}{5-p}~ {}_2F_1\big[\frac{1}{2}, \frac{4(p-5)}{(7-p)^2}, \frac{29-10p+p^2}{(7-p)^2},1 \big]  =  \frac{2\sqrt{\pi } ~\Gamma \left(\frac{29-10p+p^2}{(p-7)^2}\right)}{(5-p) ~\Gamma \left(\frac{(p-3)^2}{2 (p-7)^2}\right)}. 
\end{equation*}
Notice that there are special values of  $0\leq p\leq 6$. For $p=3$ we find the solution
\begin{equation}
 x(\tau)= \frac{1}{\sqrt{1+\tau^2}}.\label{soldpartind3}  
\end{equation}
This is the radial motion of a massive particle in AdS$_5\times S^5$. Note that $x(0)=1$ and $\dot{x}(0)=0$.

For $p=5$ the solution we wrote on the second line of eq.(\ref{solutiondpbranes}) is not well defined. To address this case, we go back to the differential equation on the first line of (\ref{solutiondpbranes}). For $p=5$ we find
\begin{align}
    4 \tanh ^{-1}\left({\sqrt{1-x^{1/2}}}\right)=\tau-\tau_0.
\end{align}
Taking the limit $x \rightarrow 1$ when $\tau \rightarrow 0$, one finds $\tau_0 = 0$. The solution is,
\begin{equation}
 x(\tau)=\frac{1}{\cosh^4(\frac{\tau}{4})} .  
\end{equation}
Finally, for $p=6$ the differential equation (\ref{solutiondpbranes}) is
\begin{eqnarray} 
& & \frac{dx}{d\tau}+ \sqrt{x(1-x^{\frac18})}=0,~~\tau=-\int \frac{dx}{\sqrt{x(1- x^{\frac18})}},\nonumber\\
& & \tau= \frac{16}{35}\left( 16+ 8x^{\frac18} + 6 x^{\frac14} +5 x^{\frac38}\right) \sqrt{1- x^{\frac18}}.
\end{eqnarray}
Notice that whilst for $p<6$, the time to reach $x=0$ is infinite, this is finite for the D6 branes \cite{Itzhaki:1998dd}. There is non-decoupling of gravity from the QFT and we leave this case aside. Please, see the word of caution below.

We can work with the implicit solution in eq.(\ref{solutiondpbranes}). To gain a better understanding, we expand the right side of  eq.\eqref{solutiondpbranes} close to the initial position $x \sim 1$. This yields 
\begin{eqnarray}
    \tau=\frac{4 \sqrt{2}\sqrt{1-x}}{(7-p)}\Bigg(1+ \frac{(153 -30 p+p^2)}{96}(1-x)+ {\cal O}(1-x)^2 \Bigg).
\end{eqnarray}
Inverting this series, we find at short times,
\begin{eqnarray}
& &  x(\tau)|_{\tau\to 0}\approx 1-\frac{(7-p)^2}{32} \tau^2 ~~\text{which implies}\nonumber\\
 & &    r(t)|_{t \sim 0}\approx r_{UV}\Big[ 1-\gamma(p) \frac{t^2}{r^2_{UV}}\Big]~;\text{with}~~\gamma(p)=\frac{(7-p)^2}{32}\Big(\frac{r_{UV}}{l} \Big)^{7-p}.
\end{eqnarray}
We can use the expressions in eqs. (\ref{complexitydpint})-(\ref{cdotdp})-(\ref{cddotdp}) to write
the complexity and its derivatives  in terms of the $x$-variable.
This gives 
\begin{eqnarray}
 & & \frac{1}{m ~r_{UV}~\sqrt{A(r_{UV})}}\left(\frac{r_{UV}}{l}\right)^{\frac{(7-p)(15-p)}{16}} ~{\cal C}_K(t)= \frac{16}{(p^2-22p+89)}\Bigg[\frac{1}{x(t)^{\frac{p^2-22p+89}{16}}} -1 \Bigg],\nonumber\\
 & & \frac{1}{m}\dot{\cal C}_K=\sqrt{\left(\frac{1}{x}\right)^{\frac{(7-p)^2}{8}} -1},\nonumber\\
 & & {\cal U}= \frac{(7-p)^2}{8 l}\Big( \frac{l}{r_{UV}}\Big)^{\frac{(p-3)^2}{16}}\Big( \frac{1}{x}\Big)^{\frac{(p-3)^2}{16}}.\label{accdp}
\end{eqnarray}
 \begin{figure}
    \centering
    \includegraphics[width=0.9\linewidth]{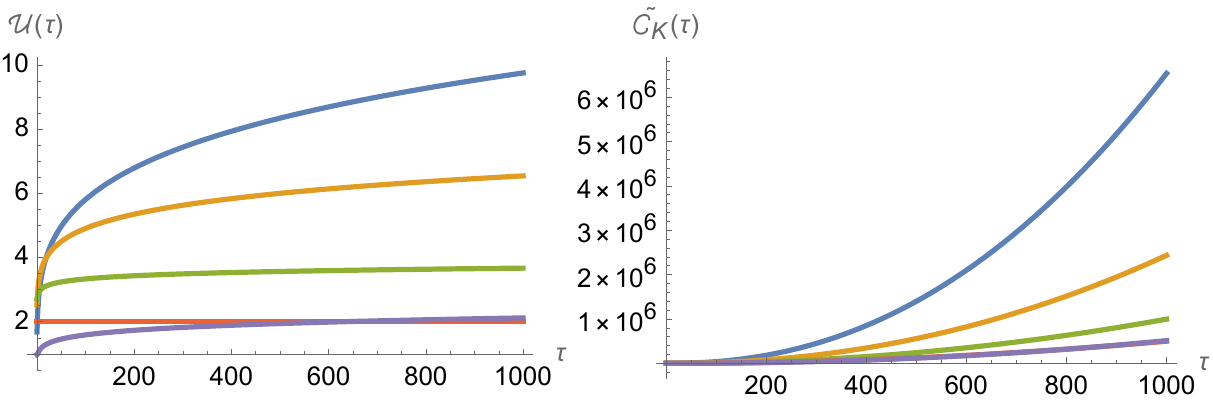}
    \caption{We plot $\mathcal{U}(\tau)$ and $\tilde{\mathcal{C}_K}(\tau)$ vs $\tau$ for different values of $p=0,1,2,3,4$, where we set $r_{UV}=10$ and $l=1$.}
    \label{figdp}
\end{figure}

In Figure \ref{figdp} we plot these quantities for different Dp-branes with $0\leq p\leq 5$.  
\subsubsection*{\underline{\bf A word of caution}}
The near horizon geometry of Dp branes, the background in eq.(\ref{bckdp})--when complemented by a Ramond $A_{p+1}$ potential--has a finite range of validity. Indeed, as explained in \cite{Itzhaki:1998dd, Boonstra:1998mp} these supergravity solutions have generically a divergent dilaton and  a divergent Ricci scalar, not necessarily at the same point. The exception is the case $p=3$.  These divergences indicate a large $r=r_{max}$ and a small $r=r_{min}$ beyond which the background is not trustable. Our calculation proceeds following \cite{Caputa:2024sux} and choosing a value $r_{UV}$ that we {\it must} take to be $r_{UV}\leq r_{max}$. As the massive probe particle falls towards the (singular) end of space (r=0), we cross the value $r_{min}$. The calculation should not be trusted below $r_{min}$. Whilst the qualitative trend of the calculated result might be correct we should take it with a grain of salt below $r_{min}$. When the technology becomes available, it would be interesting to compute both $\alpha'$ and $g_s$ corrections to the metric, dilaton and Ramond potential describing Dp branes, their effect on the radial geodesic and complexity. 
\subsubsection*{
\underline{\bf The covariant c-function and the acceleration of the complexity}
}
To close this section we point out a nice relation between the covariant central charge and the quantity ${\cal U}$, the second time derivative of the Krylov spread complexity. This new relation goes along the lines of that in eq.(\ref{firstinstance}).
The covariant c-function for Dp branes was calculated in \cite{Jokela:2026cjm}--See Section 3.2 of that paper with the exchange $d\to p+1$. We have
\begin{equation}
 c_{\text{cov}}= \left(2^{7-2p} \pi^{\frac{9-3p}{2}} \Gamma(\frac{7-p}{2})\right)^{p+1}  \frac{\text{Vol}~S_{8-p}}{G_{N}^{(10)}} \left(\frac{p}{9-p}\right)^p \left( g_{YM}^2 N\right)^{\frac{p+1}{2}} r_{UV}^{\frac{(p-3)^2}{2}} ~~x^{\frac{(p-3)^2}{2}}
\end{equation}
Comparing this with eq.(\ref{accdp}), we find
\begin{equation}
 c_{\text{cov}} \times {\cal U}^8=\Upsilon(p).   \label{expresionc-U}
\end{equation}
This constant $\Upsilon(p)$ depends on $p$, the dimensionality of the Dp brane. The expression (\ref{expresionc-U}) has a similar message as the one that eq.(\ref{firstinstance}) has. As the covariant c-function becomes smaller (that is, as we move towards  smaller values of the $r$-coordinate, the IR of the dual QFT), the second time derivative of the complexity must increase.

%\begin{align}
 %   \ddot{C}|_{t \sim 0}= m\frac{(7-p)^2}{16 l}\Big( \frac{r_{UV}}{l}\Big)^{\frac{(5-p)}{2}}\Big[1+\frac{(p-3)^2}{16}\gamma(p)t^2 \Big]+\mathcal{O}(t^4).
%\end{align}
%
%Considering $x\sim 0$ and keeping only term at leading order in the expansion we find
%\begin{align}
 %   \frac{r(t)}{l}|_{t \rightarrow \infty}=\Big( \frac{5-p}{2}\Big)^{\frac{2}{p-5}}\Big( \frac{t}{l}\Big)^{\frac{2}{p-5}}.
%\end{align}
%
%This gives
%\begin{align}
 %   \ddot{C}|_{t \rightarrow \infty }= m \frac{(7-p)^2}{16 l}\Big( \frac{r_{UV}}{l}\Big)^{\frac{(7-p)^2}{16}}\Big( \frac{5-p}{2}\Big)^{\frac{(p-3)^2}{8(5-p)}}\Big( \frac{t}{l}\Big)^{\frac{(p-3)^2}{8(5-p)}}.
%\end{align}
%
%Clearly, $p=3$ is a special case which gives $\ddot{C}/C_0=\frac{r_{UV}}{l}$ is a constant, where $C_0=\frac{m}{l}$.
%
%
%
%
In the next section, we discuss RG-flows in which the dimensionality of space-time changes. This is common when considering compactifications of QFTs (or CFTs in the examples below) in $d$-dimensions, on finite volume $q$-dimensional manifolds $\Sigma_q$. For energies below the inverse scale set by the compactification manifold $\Sigma_q$, the theory ceases to explore $\Sigma_q$ and the low energy modes probe only $(d-q)$-dimensions, realising a spontaneous compactifications of the CFT or QFT.

\section{RG flows across dimensions}\label{sectionRGflowacross}
We consider here two examples of spontaneous compactification of CFTs (in dimension four and six is respectively), that after an RG-flow is triggered by compactification of a hyperbolic two-manifold, flow into an IR description in two less dimensions, reaching at low energies CFTs in dimensions two and four respectively.

The first example is obtained using the contents of \cite{Buchel:2006gb,Gauntlett:2007ma, Gauntlett:2009zw, Donos:2014eua, Donos:2008ug}. In fact, in \cite{Bea:2015fja}, compactifications of the ${\cal N}=1$ Klebanov-Witten SCFT  theory on two dimensional hyperbolic manifolds were found. We are interested on a solution first presented in \cite{Gauntlett:2006af} and further studied in \cite{Bea:2015fja}. The background contains a metric and a five form. The dilaton, axion,  Ramond and Neveu-Schwarz two forms vanish. In this paper, we only need the metric (for details see Section 2 of \cite{Bea:2015fja}), that reads
\begin{eqnarray}
& &    \frac{ds^2}{l^2}=\frac{e^{3r}}{\sqrt{1+e^{2r}}}(-dt^2+dx^2)+dr^2+\frac{1+e^{2r}}{3}(d\alpha^2+\sinh^2\alpha d\beta^2)+d\hat{s}^2_{T^{1,1}}\label{DGKmetric}\\
& & \text{where},\nonumber\\
&    &d\hat{s}^2_{T^{1,1}}=\frac{1}{6}d\Omega^2_2(\theta_1 , \phi_1)+\frac{1}{6}d\Omega^2_2(\theta_2 , \phi_2)+\frac{1}{9}\Big(d\psi+\cos\theta_1 d\phi_1+\cos\theta_2 d\phi_2+\mathcal{A}_1 \Big)^2\nonumber\\
 &   &\mathcal{A}_1=\cosh\alpha d \beta.\nonumber
\end{eqnarray}
This geometry interpolates between  AdS$_5\times \tilde{T}^{1,1}$ (for $r\to\infty$), holographically dual to a twisted 4d CFT that upon compactification on the finite-cell of a hyperbolic plane flows in the IR ($r\to-\infty$) to AdS$_3\times M_7$, dual to a 2d SCFT preserving two Poincare supercharges. The details of this twisted compactification, like the operator VEV that triggers the RG-flow and other observables are described in Section 6 of \cite{Bea:2015fja}.

\begin{figure}
    \centering
    \includegraphics[width=0.9\linewidth]{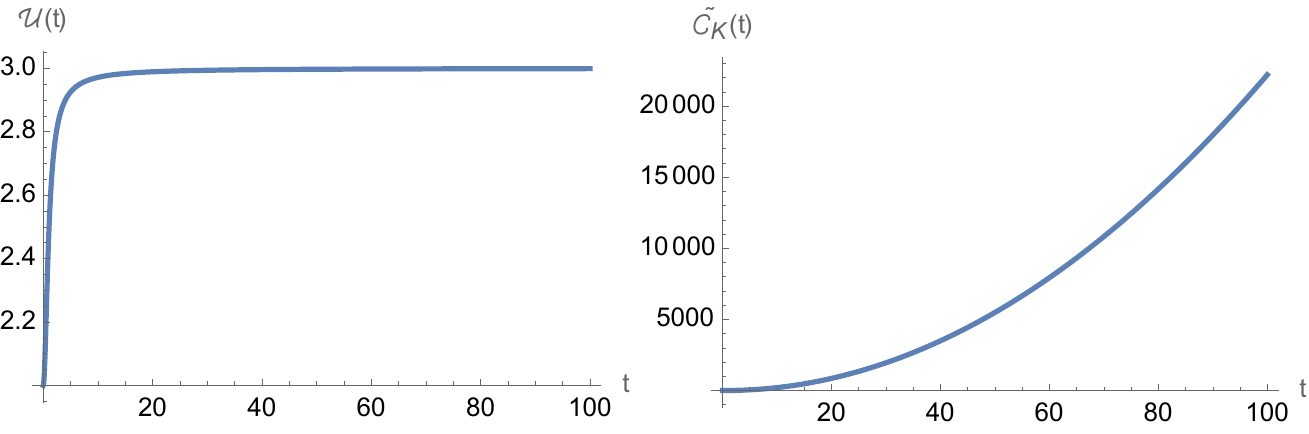}
    \caption{Plot of second derivative of complexity $\mathcal{U}(t)$ and the rescaled complexity $\tilde{\mathcal{C}}_K(t)$ vs $t$.}
    \label{fig3}
\end{figure}
Comparing  with eqs.(\ref{genericbackground})-(\ref{specialxx}), we identify the functions
\begin{equation}
    A(r)=\frac{e^{3r}}{\sqrt{1+e^{2r}}}~;~B(r)=\frac{1}{A(r)}.
\end{equation}
We are in the situation described by eq.(\ref{formulasutiles}). We find for the complexity and its derivatives,
\begin{eqnarray}
& & \frac{3}{m \sqrt{A(r_{UV})}}{\cal C}_K(t)= e^{-3 r(t)}\left( 1+ e^{2r(t)}\right)^{\frac{3}{2}} - e^{-3 r_{UV}}\left( 1+ e^{2r_{UV}}\right)^{\frac{3}{2}},\label{complexityDG}\\
  & & \frac{1}{m} \dot{\cal C}_K(t)=
  \sqrt{   \sqrt{\frac{e^{2 r(t)}+1}{{e^{2 r_{UV}}+1}} }~~ e^{3 \left( r_{UV}- r(t)\right)} -1  }
  \label{complexitydotDG}\\
&    & \frac{2}{m\sqrt{A(r_{UV})}}\ddot{\cal C}_K(t)={\cal U}(t)=\frac{3+ 2 e^{2r(t)}}{1+ e^{2r(t)}}.\label{complexitydotdotDG}
\end{eqnarray}

See Fig.\ref{fig3} for a variation of the above entities (\ref{complexityDG}) and (\ref{complexitydotdotDG}) with time $t$, where $\tilde{\mathcal{C}}_K(t)$ is the entity on the l.h.s. of (\ref{complexityDG}).

The function $r(t)$ is obtained by integrating eq.(\ref{eq:rdot_Krylov}),
\begin{eqnarray}
& & t-t_0= -\int_{r_{UV}}^{r(t)}\frac{dz}{ \sqrt{ \frac{e^{3z}}{\sqrt{1+ e^{2z}}} \Bigg(1- \sqrt{\frac{e^{2 r_{UV} }+1}{{e^{2 z}+1}} }~~ e^{3 \left(z- r_{UV}\right)} \Bigg)  }}    
\end{eqnarray}
The covariant c-function for this background was calculated in Section 8.2 of \cite{Bea:2015fja}. The result is
\begin{equation}
c_{\text{cov}}= \frac{\widehat{\cal N}}{G_{N}^{(10)}} \Bigg( \frac{1+e^{2r}}{1+ 2 e^{2r}}\Bigg)^3.   \label{cuti} 
\end{equation}\\
Here, $\widehat{\mathcal{N}}$ involves the volume of the $T^{1,1}$ and the volume of the hyperbolic plane (proportional to the genus of the Riemann surface.

A relation between $c_{\text{cov}}$ and the (scaled) acceleration of the complexity ${\cal U}(t)$ in eq.(\ref{complexitydotdotDG}) can be found. This relation reads
\begin{equation}
c_{\text{cov}}=  \frac{\widehat{\cal N}}{G_{N}^{(10)}} \times \frac{1}{(4-{\cal U})^3}. \label{enzo}
\end{equation}\\

There is a subtlety in this example. This subtlety is rooted on the fact that the RG flow breaks the original Lorentz $SO(1,3)$ of the dual 4d SCFT into $SO(1,1)$ the Lorentz group for a 2d SCFT. Let us explain this in some detail.

The expression in eq.(\ref{enzo})  shows a relation between $c_{\text{cov}}$ and $\cal U$. When the acceleration of the complexity grows, the particle is flowing towards the IR--see eq.(\ref{complexitydotdotDG}). Close to the fixed points, the quantity ${\cal U}$ takes fixed values. 
Note that far in the UV (taking $r_{UV}\to\infty$), deep in the AdS$_5$ region we have ${\cal U}_{UV}=2$. On the other end of the energy flow, for $r\to-\infty$, we find ${\cal U}_{IR}=3$. The acceleration of the complexity grows towards the IR and stabilises for very short and large times, for which the particle is close to fixed points, corresponding to the UV and the IR of the RG-flow QFT. 
 The c-function is monotonic, but {\it grows} when moving towards the IR. In fact, from eq.(\ref{cuti}) $c_{\text{cov}}(+\infty)=\frac{\widehat{\cal N}}{8 G_N^{(10)}}$ and $c_{\text{cov}}(-\infty)= \frac{\widehat{\cal N}}{G_N^{(10)}})$.  This is a consequence of the breaking of Lorentz invariance in the 4d SCFT (precluding the usual picture of c-function decreasing towards the IR) and the fact that massless modes exist on the hyperbolic space that have an effect in the number of degrees of freedom in the IR. Hence, the c-function and the acceleration of the complexity are {\it co-monotonic}.

We now study a different example, corresponding to a flow from the ${\cal N}=(0,2)$ SCFT in six dimensions to four dimensional SCFTs with eight Poincare supercharges. The flow is the one in \cite{Maldacena:2000mw}.
\subsection{RG flow of wrapped  M5 branes}
Consider the ${\cal N}=(0,2)$ six dimensional SCFT describing the decoupled low-energy dynamics of a stack of M5 branes. We compactify this SCFT on a two dimensional hyperbolic space, with a topological twist. The hyperbolic space can have punctures, but the example discussed below is the simplest in this class. Flowing to the IR gives an infinite family of ${\cal N}=2$ SCFTs in four dimensions, this is studied in  more generality in \cite{Gaiotto:2009gz, Gaiotto:2009we, Lin:2004nb, Aharony:2012tz, Reid-Edwards:2010vpm, Lozano:2016kum, Nunez:2019gbg, Nunez:2018qcj}. We consider the simplest of all such solutions described in Sections 4 and 7.3 of \cite{Maldacena:2000mw}. 

The configuration of wrapped M5 branes is described by a metric and a four form, solution to the equations of motion of eleven dimensional supergravity.
We quote only the metric, that reads,
\begin{eqnarray}
  & &   ds^2_{11}=\tilde{\Delta}^{1/3}\Bigg[ ds^2_7+\frac{1}{ \tilde{\Delta}}\Big( \mathbb{A}d\theta^2+\mathbb{B}d\psi^2+\mathbb{C} d\theta d\psi +\frac{\mu_1^2}{X_1}(d\phi_1 - A_1)^2+\frac{\mu_2^2}{X_2}(d\phi_2 - A_2)^2\Big)\Bigg],\label{Mmetric}\\
    & & ds^2_7=e^{2f(r)}(-dt^2+dx^2_1+dx^2_2+dx^2_3+dr^2)+\frac{e^{2g(r)}}{y^2}(dx^2+dy^2).\label{7metric}
\end{eqnarray}
We have defined the functions 
\begin{eqnarray}
& & \mathbb{A}=\sin^2\theta \Big[ \frac{\cos^2\psi}{X_2}+\frac{\sin^2\psi}{X_1}\Big]+\frac{\cos^2\theta}{X_0},~~~\mathbb{B}=\cos^2\theta \Big[\frac{\sin^2\psi}{X_2}+\frac{\cos^2\psi}{X_1} \Big],\nonumber\\
& &\mathbb{C}=2\sin \theta \cos\theta \sin\psi \cos \psi \Big[\frac{1}{X_1}-\frac{1}{X_2} \Big],~~ \tilde{\Delta}=X_0 \sin^2\theta + \cos^2\theta\left(X_1\sin^2\psi+ X_2 \cos^2\psi \right),  \nonumber\\
& &A_1=\frac{a}{y}dx ~;~A_2=\frac{b}{y}dx, ~~X_0(r), ~X_1(r), X_2(r).\label{M5functions}
\end{eqnarray}
Notice that the background is of the generic form of eq.(\ref{genericbackground}), with warp factors that depend both on the radial coordinate but also on the coordinates of the internal space, in this case ($\theta,\psi$)\footnote{This requires caution when proposing a geodesic motion $r(t)$ for a massive probe, as there may be coupling with the internal angles.}.

Depending on the parameters $(a,b)$ one can write different BPS equations that describe SUSY solutions of eleven dimensional supergravity, see \cite{Maldacena:2000mw, Lin:2004nb} for details. In fact, for $(4a=1, b=0)$
an explicit solution of these BPS equations exist. It reads,
\begin{eqnarray}
    & & X_0=X_2=e^{2\lambda(\rho)}, ~~X_1=e^{-3\lambda(\rho)},~~4a=1,~~b=0,\label{solutionmtheory}\\
  & &   e^{5\lambda(\rho)}=\frac{4e^{2\rho}+2 
  }{4e^{2\rho}+1}, ~~e^{2g(\rho)}= e^{\lambda(\rho)}\left( e^{2\rho}+\frac14\right),~~e^{2f(\rho)}= e^{2\rho+ \lambda(\rho)},~~e^{f(r)}dr= e^{-2\lambda(\rho)}d\rho.\nonumber
\end{eqnarray}
Notice that for $\rho\to+\infty$ (the UV of the QFT), we find $\lambda\approx 0$ which implies $X_1\sim X_2\sim X_0\sim 1$ and  $e^{2g}\sim e^{2f}\sim e^{2\rho}$. The space time asymptotes to AdS$_7\times S^4$. This represent the six dimensional SCFT on $R^{1,3}\times H_2$.
On the other hand, for $\rho\to-\infty$, we find $e^{5\lambda}=2$, $4e^{2g}=e^{\lambda}$, $e^{2f}\sim e^{2\rho}$, $X_{1}\sim X_{2}\sim X_{0}\sim \text{constant}$. The geometry asymptotes to $AdS_5\times M_6$. This represents a four dimensional SCFT, with eight Poincare SUSYs and $SU(2)\times U(1)$ global R-symmetry.

To study the complexity we consider a geodesic falling from a  certain fixed UV position, and with zero initial velocity. In general, we propose a configuration in which {\it all coordinates} depend on $t$, that is, $r(t),x_i(t),y(t), x(t),\theta(t),\psi(t), \phi_1(t),\phi_{2}(t)$. We study the equations of motion of this coupled system and find that for constant values of $[x_{1},x_2,x_3, x,y,\phi_{1},\phi_2]$ their equations of motion are solved. We remain with a system of equations for three functions [$r(t),\theta(t),\psi(t)$].
This system can be further truncated by choosing certain particular constant values of $(\theta,\psi)$. The equations of motion for the $\theta,\psi$-coordinates become 
\begin{equation}
    \partial_\theta \tilde{\Delta}= \partial_\psi \tilde{\Delta}=0.\label{4.1.3}
    \end{equation}
    The specific value of these coordinates has an influence on the dynamics. In fact, there are various allowed set of values that solve eq.(\ref{4.1.3}) automatically, the equations of motion for the $(\theta(t),\psi(t))$ coordinates. We take as representative the values
\begin{equation}
 (\theta(t),\psi(t))=(\theta_0,\psi_0)= (0,0)~\longrightarrow \tilde{\Delta}= X_2=e^{2\lambda}. %(0,\frac{\pi}{2});~ (\frac\pi2, 0); ~(\frac{\pi}{2},\frac\pi2).   
\end{equation}
The choice of different constant values for the angles changes the value of $\tilde{\Delta}(r,\theta_0,\psi_0)$. The induced metric for the probe particle falling under geodesic motion and the associated Lagrangian for the radial coordinate $r$ in this consistently truncated configuration reads
\begin{eqnarray}
& &       ds^2=-A(r)dt^2+A(r)B(r)dr^2, ~~
    A(r)=\tilde{\Delta}^{1/3}(r)e^{2f(r)}~;~B(r)=1,\nonumber\\
    & &  L=-m\sqrt{\tilde{\Delta}^{1/3}e^{2f(r)}-\tilde{\Delta}^{1/3}e^{2f(r)}\dot{r}^2}.\label{lagrangianomtheory}
\end{eqnarray}
Let us appreciate an interesting point that eq.(\ref{lagrangianomtheory}) is making.
The factor $\tilde{\Delta}^{\frac13}$ shows explicitly that a massive probe moving radially in the full eleven-
dimensional geometry is sensitive to its position on the internal $S^4$. A radial geodesic
computed solely in the seven-dimensional metric, eq.(\ref{7metric}) would omit this factor and need not
agree with the eleven-dimensional trajectory. It is therefore necessary both to select a
stationary internal point and to state that choice before applying the one-dimensional
complexity formulae.
\begin{figure}
    \centering
    \includegraphics[width=0.9\linewidth]{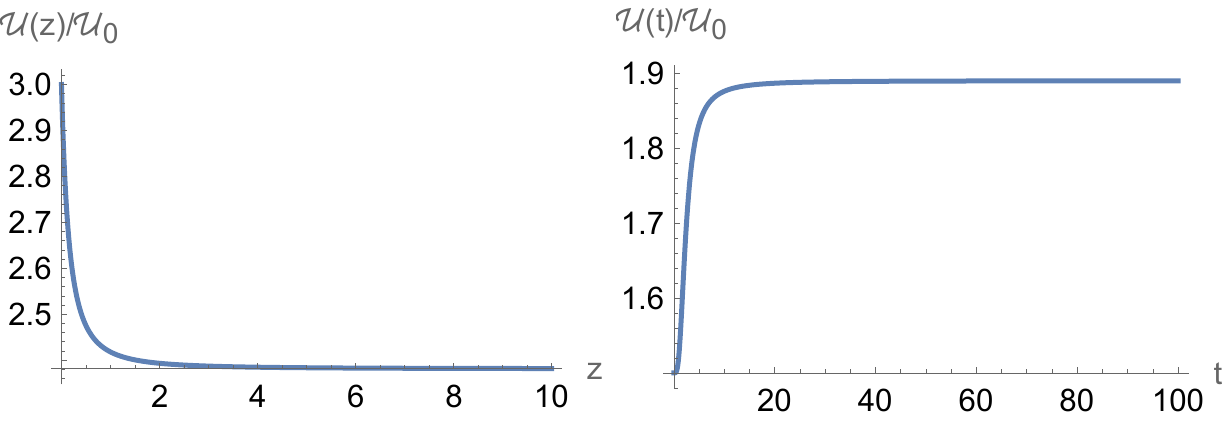}
    \caption{We plot $\mathcal{U}/\mathcal{U}_0$ against the radial coordinate $z$ and time $t$.}
    \label{figU}
\end{figure}

%
%The presence of the factor $\tilde{\Delta}(r)$ in the lagrangian is telling of the influence of the internal space  on the radial geodesic. In other words, had we computed the radial trajectory in a seven dimensional truncation of M-theory--the metric of such truncation is that of eq.(\ref{7metric})-- we would have not found the factor of $\tilde{\Delta}(r)$. 
%In what follows we choose the particular solution
%\begin{equation}
% \theta_0=n\pi, ~~\psi_0=0,~~~\tilde{\Delta}= e^{2\lambda}.   
%\end{equation}
%Other choices can produce a different result $\tilde{\Delta}$.

Transforming to the coordinate $\rho$, where $e^{f}dr= e^{-2\lambda} d\rho$ as indicated in eq.(\ref{solutionmtheory}). This implies $e^{2f(r)}\dot{r}^2=e^{-4\lambda(\rho)}\dot{\rho}^2$. Putting all together, we have for the Lagrangian in eq.(\ref{lagrangianomtheory})
\begin{eqnarray}
& &
    L=-m\sqrt{A(\rho)\left(1-B(\rho)\dot{\rho}^2\right)},\label{lagrangianmtheoryfinal}\\
& &A= \tilde{\Delta}^{\frac{1}{3}} e^{2f(\rho)},~~~ B= e^{-2f(\rho) -4\lambda(\rho)},\nonumber\\
& & A(\rho)=2^{1/3}e^{2 \rho } \Big[\frac{1+2e^{2\rho}}{1+4 e^{2 \rho }}\Big]^{1/3}~;~B(\rho)=\frac{e^{-2 \rho }}{2}\Big[\frac{1+4e^{2\rho}}{1+2 e^{2 \rho }}\Big].\nonumber
\end{eqnarray}
We use eqs.(\ref{cdot}),(\ref{cdotdot}),(\ref{complexityrt}) to write,
\begin{eqnarray}
& &\frac{{\cal C}_K(t)}{m \sqrt{A(\rho_{UV})}}= \frac{1}{2^{2/3}}\int_{\rho(t)}^{\rho_{UV}} d\rho~ e^{-2\rho}
\left( \frac{1+4e^{2\rho}}{1+ 2 e^{2\rho}}\right)^{\frac23} \label{complexitym5}\\
& & \frac{\dot{\cal C}_K(t)}{m}= \sqrt{ e^{2\rho_{UV}-2\rho} \Bigg(\frac{\left(1+2 e^{2\rho_{UV}} \right) \left( 1+ 4 e^{2\rho}  \right)}{\left(   1+ 4 e^{2\rho_{UV}}\right) \left( 1+2 e^{2\rho}  \right)} \Bigg)^{\frac13} -1           }\label{cdotm5}\\
& & 2\frac{\ddot{\cal C}_K(t)}{m\sqrt{A(\rho_{UV})}} ={\cal U}= %\frac{A'(\rho)}{\sqrt{A^3(\rho)B(\rho)}}=
\frac{2^{\frac43}}{3}\frac{  \left(16 e^{2 \rho }+24 e^{4 \rho }+3\right)}{ \Big(\left(4 e^{2 \rho }+1\right)^2 \left({2 e^{2 \rho }+1}\right)\Big)^{\frac23}}.\label{Um5}
\end{eqnarray}\\
All these should be evaluated using $\rho(t)$ obtained from,
\begin{equation}
 t-t_0=\int_{\rho}^{\rho_{UV}} d\rho~~  \sqrt{\frac{(1+ 4 e^{2\rho})}{2 e^{2\rho}(1+ 2 e^{2\rho}) 
    \Bigg( 1- e^{2\rho-2\rho_{UV}} \big(\frac{\left(1+2 e^{2\rho }\right) \left( 1+ 4 e^{2\rho_{UV}}  \right)}{\left(   1+ 4 e^{2\rho} \right ) \left( 1+2 e^{2\rho_{UV}}  \right)} \big)^{\frac13}\Bigg)   }   } 
\end{equation}
Similarly, one can compute the c-function along the flow.
This gives the same result if done in eleven dimensions or in seven dimension (that is the dimensionally of the gauged supergravity solution leading after lift to the eleven dimensional background used here). One finds,
\begin{equation}
c_{\text{cov}}=\left( \frac{5}{2}\right)^{5}\frac{\tilde{\cal N}}{G_{N}^{(11)}}  \Bigg[ \frac{e^{-2\lambda}}{3f'+2g'}\Bigg]^5.\label{ccovm5}  
\end{equation}
We  denoted with primes the derivative respect to $\rho$ and $\tilde{\cal N}$ denotes the product of the volumes of the four-sphere and the hyperbolic plane. 

\begin{figure}
    \centering
    \includegraphics[width=0.8\linewidth]{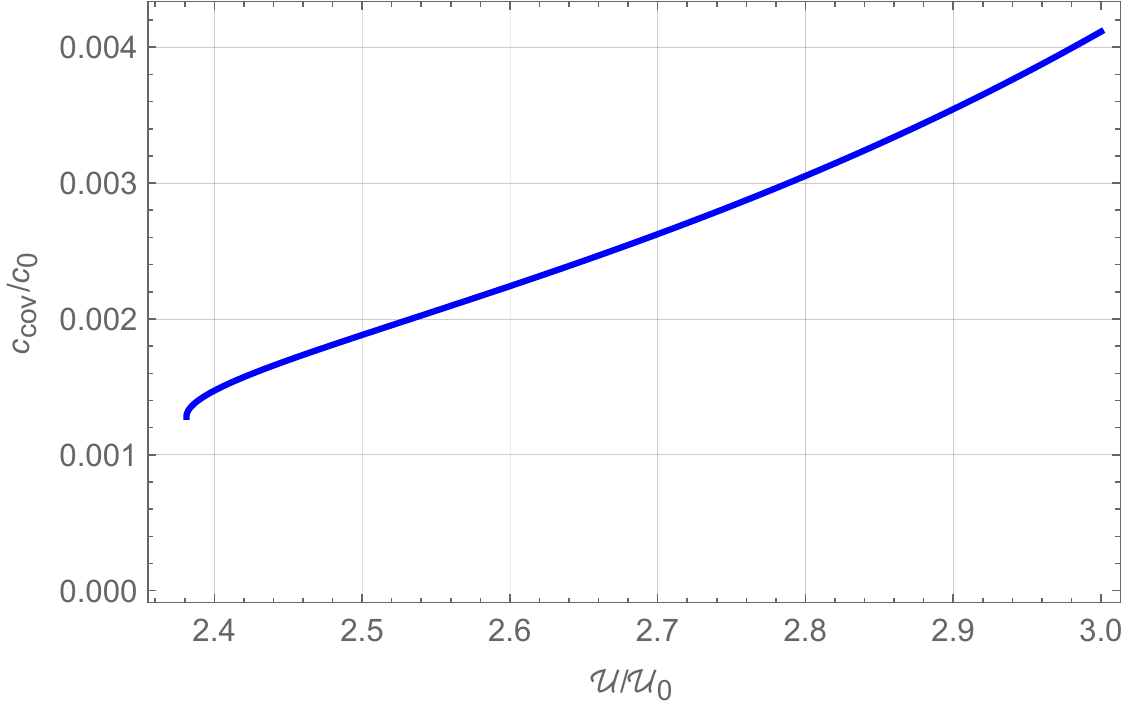}
    \caption{Plot for $c_{cov}/c_0$ vs. $\mathcal{U}(z)/\mathcal{U}_0$.}
    \label{fig:placeholder4}
\end{figure}
Writing $c_{\text{cov}}$ in terms of $\cal U$ is not as  analytically explicit as in previous cases. We can offer the following 'parametric' representation, in terms of $z=e^{2\rho}$:
 \begin{eqnarray}
& & c_{\text{cov}}=     \left( \frac{5}{2}\right)^{5}\frac{\tilde{\cal N}}{4 G_{N}^{(11)}}  \Bigg[ \frac{(2z+1)^3 (4z+1)^7}{\Big( 3+ 24 z+ 40 z^2  \Big)^5}\Bigg],~~
{\cal U}= \frac{2^{\frac43}}{3} \frac{24z^2+16 z+3}{\Big( (4z+1)^2(2z+1) \Big)^{\frac23}}.\label{julianal}\\
& & \text{Where we defined}~z= e^{2\rho}.\nonumber
 \end{eqnarray}\\
We can expand these expressions for large $z$ (the UV of the QFT) and $z\sim 0$ (the IR of the QFT). In the UV we find
\begin{eqnarray}
& & c_{\text{cov}}(z)\approx \frac{c_0}{3125}(1+\frac{1}{4z}+....),~~{\cal U}(z)\approx \frac{3{\cal U}_0}{2^{\frac13}} (1+ \frac{1}{36z^2}+....),\label{dibu}\\
& &\text{where}~~c_0= \left( \frac{5}{2}\right)^{5}\frac{\tilde{\cal N}}{G_{N}^{(11)}}, ~~~{\cal U}_0= \frac{2^{\frac43}}{3}.\nonumber
\end{eqnarray}
Solving for $z(U)$ and plugging in $c_{\text{cov}}$, we find
\begin{equation}
c_{\text{cov}}({\cal U})\approx \frac{c_0}{4\times 3125}\left( 4 +2^{\frac23}\sqrt{6\left( \frac{\cal U}{\cal U}_0 -\frac{3}{2^{\frac13}}\right)} +....  \right)  .
\end{equation}
On the other hand, for $z\to 0$, the IR of the SQFT we find,
\begin{equation}
c_{\text{cov}}(z)\approx \frac{c_0}{4\times 243}(1-6z+....), ~~~{\cal U}(z)\approx3 {\cal U}_0 (1-\frac{4z}{3}+....) . 
\end{equation}
This implies that for $z\to 0$, we find
\begin{equation}
c_{\text{cov}}({\cal U})\approx \frac{c_0}{4\times 243}(\frac32 \frac{\cal U}{\cal U}_0 -\frac72)    
\end{equation}
We parametrically plot $c_{\text{cov}}$ in terms ${\cal U}$ for $0\leq z\leq \infty$. See Figure \ref{fig:placeholder4},
%\textcolor{blue}{In the IR, we have the following expansions
%\begin{align}
% & \frac{c_{cov}}{c_0}|_{z \sim 0}=\bar{c}(z)=\frac{1}{243}-\frac{2 z}{81}+\frac{148 z^2}{729}\\
% & \frac{\mathcal{U}(z)}{\mathcal{U}_0}|_{z \sim 0}=\bar{\mathcal{U}}(z)=3-4 z+20 z^2.
%\end{align}
%Upon solving (5.24) we find $z=z(\bar{\mathcal{U}})$ which after substituting into (5.23) yields $\bar{c}(z)-\bar{c}_0 \sim \bar{\mathcal{U}}(z)$.
%On the other hand, in the UV we find
%\begin{align}
 %& \frac{c_{cov}}{c_0}|_{z \rightarrow \infty}=\hat{c}(z)=\frac{4}{3125}+\frac{1}{3125 }\frac{1}{z}-\frac{3}{62500 }\frac{1}{z^2}+\mathcal{O}(z^{-3})\\
% & \frac{\mathcal{U}(z)}{\mathcal{U}_0}|_{z \rightarrow \infty}=\hat{\mathcal{U}}(z)=\frac{3}{\sqrt[3]{2}}+\frac{1}{12 \sqrt[3]{2} }\frac{1}{z^2}+\mathcal{O}(z^{-3}).
%\end{align} 
%Like before we solve $z=z(\hat{\mathcal{U}})$, which finally yields $\hat{c}(z)-\hat{c}_0 \sim \hat{\mathcal{U}}(z)$.}
%Similarly to what happens in eq.(\ref{cuti}), this monotonic quantity reaches
%a value in the UV ($\rho\to\infty$) that is smaller than the IR value ($\rho\to-\infty$). 
the same analysis as that around eq.(\ref{enzo}) applies:
for a release point sufficiently far in the UV, ${\cal U}(t)$ begins close to its AdS$_7$ value  ${\cal U}_{UV}=2$.
As the particle falls towards smaller $\rho$, $ {\cal U}(t)$ increases monotonically and approaches
${\cal U}_{IR}=2^\frac{4}{3}$ at late times (Fig.\ref{figU}), when the geometry approaches its AdS5 region. The covariant
c-function is {\it co-monotonic}: it also increases towards the IR, with 
\begin{equation}
\frac{c_{\text{cov}[}UV]}{c_{\text{cov}}[IR]}=\frac{972}{3125}.
\end{equation}
Thus this flow exhibits a direct, rather than inverse, correlation between
the acceleration of spread complexity and the covariant c-function. The increase of
$c_{\text{cov}}$ is compatible with the flow-across-dimensions interpretation because the hyperbolic
compactification breaks six-dimensional Lorentz invariance and admits massless modes
in the lower-dimensional description. As a follow-up to this, it would be nice to study other RG-flows between fixed points across dimensions, like those in \cite{Nunez:2008wi, Nunez:2023nnl, Nunez:2023xgl, Gursoy:2002tx, Nunez:2001pt, Legramandi:2021uds}.
\subsection{{Comparing the RG flows}}
For RG-flows that preserve the dimensionality of the space time the covariant c-function and the (normalised) acceleration of the spread complexity are controled by one and the same scale. The energy condition implying the monotonic decrease of $c_{\text{cov}}$ ensures the monotonic increase of ${\cal U}$. The expressions $c_{\text{cov}}\times \left({\cal U} \right)^{d-1}\sim \text{constant}$ and $c_{\text{cov}}\times \left({\cal U} \right)^{8}\sim \text{constant}$ are examples of this 'inverse' relation.

The compactification qualitatively changes this picture. The compact manifold introduces a second scale (in the cases studied here, the size of $H_2$). The generic expression for $c_{\text{cov}}$, for a background of the form
\begin{equation}
ds^2\sim e^{2a(r)}dx_{1,d-q-1}^2+ e^{2b(r)}d\Sigma_q +g_{ij}(dy_i+ A^i)(dy_j+A^j).
\end{equation}
is of the form (see\cite{Bea:2015fja}) $c_{\text{cov}}\times (a'+b')^{d-1}\sim \text{constant}$. This produces a relation that is not necessarily inverse. The effect of the second scale appear in the term $b'(r)$.

In fact, using eq.(\ref{enzo}) and  defining
\begin{equation}
{\cal Q}=     \left(\frac{\widehat{\cal N}}{G_{N}^{(10)} c_{\text{cov}}}\right)^{\frac13} .
\end{equation}
We have the 'conservation law'
\begin{equation}
  {\cal Q}+ {\cal U}=4.  
\end{equation}
Field theoretically, the spectrum of the compactified QFT contains massive KK modes that decouple at low energies, and also zero modes (these are the effect of the twisting procedure). One can imagine a Krylov chain with two main sectors: one representing the hopping between KK-modes, another the hopping between zero modes and finally an interaction between them. It would be nice to formalise these ideas.

Let us now present some summary and conclusions.

\section{Conclusions and closing comments}
In this paper we have investigated spread Krylov complexity along holographic
renormalisation-group flows and compared its second time derivative with a covariant
holographic \(c\)-function.  The analysis was motivated by the proper-momentum
prescription of \cite{Caputa:2024sux, Fan:2024iop, He:2024pox, Li:2025fqz, Li:2026pdh}, according to which the rate of spread
complexity of a locally excited state is measured by the proper radial momentum of a
falling bulk probe.  This identification makes the complexity calculable in geometries
where the full boundary survival amplitude and Lanczos sequence are not presently known.
It also provides a scale-resolved probe: as the particle falls, its radial position samples
successively lower energy regions of the dual field theory.

Our main conceptual point was to focus not only on \({\cal C}_K\) or its growth rate, but
on the normalised acceleration
\[
 {\cal U}(t)=\frac{2}{m\sqrt{A(r_{\rm UV})}}\,\ddot{\cal C}_K(t).
\]
For a general radial metric this quantity is a local combination of the redshift and
radial metric functions.  It can therefore be compared directly with geometrically
defined central functions.  We have found that this comparison produces simple and
non-trivial relations in every class of examples considered.  This is the principal new
result of the work: the covariant \(c\)-function, which measures an effective density of
degrees of freedom, is related to the acceleration with which a state spreads along its
Krylov chain.

For Lorentz-invariant domain-wall flows in a fixed spacetime dimension, the relation is
particularly transparent,
\[
 c_{\rm cov}=\frac{1}{G_N^{(d+1)}{\cal U}^{d-1}}.
\]
The positivity condition on the scalar kinetic metric implies \(a''(r)\leq0\).  It follows
simultaneously that \(c_{\rm cov}\) decreases towards the infrared and that \({\cal U}\)
increases with the infall time.  The two observables are therefore anti-correlated.  The
GPPZ example illustrates this mechanism explicitly: the effective number of degrees of
freedom decreases as the acceleration of spread becomes larger, although quantitative
statements sufficiently close to the singular endpoint must be treated with caution.

The D\(p\)-brane backgrounds provide a genuinely top-down and non-conformal extension.
Despite their different dimensionalities and radial scalings, the family obeys
\[
 c_{\rm cov}~{\cal U}^{8}=\Upsilon(p).
\]
The \(p\)-dependence is carried by the constant \(\Upsilon(p)\), whereas the exponent is
uniform across the family.  The conformal D3 case is distinguished by a constant
\({\cal U}\).  For \(p\neq3\), the result exhibits explicitly how generalised conformal
structure organises both the central function and the Krylov acceleration.  Because the
supergravity description has a finite window of validity, the portions of the trajectory
beyond the dilaton or curvature bounds should be viewed as qualitative extrapolations.

A qualitatively different pattern emerged for twisted compactifications and flows across
dimensions.  In the flows from four to two dimensions and from six to four dimensions,
\(c_{\rm cov}\) and \({\cal U}\) are co-monotonic.  We obtained an explicit algebraic
relation for the compactified Klebanov--Witten theory and a parametric relation, with
ultraviolet and infrared expansions, for the wrapped-M5 solution.  The contrast with
same-dimensional domain walls is not a violation of a conventional \(c\)-theorem.  The
compactification changes the effective Lorentz group and the dimension in which the
infrared degrees of freedom propagate; protected lower-dimensional modes and the
topological twist reorganise the spectrum.  Central charge data in different spacetime
dimensions are not compared by the same monotonicity theorem or even by quantities with identical
normalisation.  In this setting an increasing covariant central function can accompany an
increasing complexity acceleration.  We regard this reversal from anti-correlation to
co-monotonicity as evidence that \({\cal U}\) distinguishes depletion of degrees of
freedom from their reorganisation into lower-dimensional sectors.

The result should be interpreted with an important qualification.  The
proper-momentum/spread-complexity equality is established sharply for a particular class
of AdS$_3$/CFT$_2$ states \cite{Caputa:2024sux}; its use in more general top-down,
non-conformal backgrounds is a physically motivated extension.  Likewise, \({\cal U}\)
is not itself a Lanczos coefficient.  It is the acceleration of the mean position of the
Krylov wavefunction and hence constrains an aggregate probability current on the chain.
Different spectral measures can share the same mean position over a limited time
interval.  A complete field-theory derivation must therefore reconstruct, or otherwise
constrain, the survival amplitude and the Lanczos sequence associated with the bulk
probe.

Several developments could make this connection more precise.  First, one may expand the
bulk trajectory and \({\cal C}_K(t)\) at short times and match the coefficients to spectral
moments.  This would determine the first Lanczos data and test whether the combinations
selected by \(c_{\rm cov}\) possess a universal large-\(N\) interpretation; the recursive
methods of Refs.~\cite{Qu:2025lgo,Muck:2026top} are particularly useful for this purpose, see also Appendix \ref{appendixa}.
Second, compactification flows naturally suggest a block-Lanczos or
symmetry-resolved description in which the lower-dimensional zero modes and massive
Kaluza--Klein towers define coupled Krylov sectors
\cite{Caputa:2025mii,Caputa:2025ozd,Craps:2024suj}.  In such a formulation,
co-monotonicity could arise from probability transfer between blocks even when the
radial propagation within each block resembles that of a same-dimensional flow.
Third, it would be valuable to compare the mean position with the variance, Krylov
entropy and higher cumulants.  Also, it might be instructive to study the relation between ${\cal U}$ and $c_{\text{cov}}$ in confining models, like those of \cite{Anabalon:2021tua, Anabalon:2024che, Chatzis:2024kdu, Chatzis:2024top, Chatzis:2025hek, Fatemiabhari:2024aua}.

It would also be interesting to determine whether the inverse and co-monotonic relations
found here admit a more general theorem.  A plausible formulation would combine an energy
condition for the bulk geometry, monotonicity of the relevant covariant central function,
and a consistent one-dimensional truncation of the probe dynamics.  Same-dimensional
domain walls and compactification flows would then correspond to different signs in the
relation between the radial derivative of the central function and the divergence of the
Krylov probability current.  Establishing such a result directly in field theory would
provide a new bridge between RG irreversibility, spectral recursion and quantum
information spreading.  The examples analysed in this work supply concrete evidence that
such a bridge exists and that the acceleration of spread complexity contains information
not captured by the complexity or the central function separately.

\section*{Acknowledgements} We want to thank various colleagues for their input that improved the contents and presentation of this work. In particular we thank: Dimitrios Chatzis, Ali Fatemiabhari, Horatiu Nastase, Alfonso V. Ramallo, Javier Subils. CN is supported by  STFC’s grants UKRI4243, ST/Y509644- 1, ST/X000648/1 and ST/T000813/1. DR acknowledges the Mathematical Research Impact Centric Support (MATRICS) grant (MTR/2023/000005) received from ANRF, India.

\appendix
\section{Comments on Lanczos coefficients}\label{appendixa}
The goal of this very short appendix is to write some comments and formulas that could be applied to the examples studied in this work. We leave this for a more dedicated work. The goal is to use holography where it is a trustable calculation-tool to learn things about the microscopic Krylov chain. As we have strongly coupled systems, it is very hard to perform trustable calculations in the QFT side (unless a lattice formulation is available).

First, we start with the Ehrenfest theorem for the spread complexity derived  and further elaborated in \cite{Erdmenger:2023wjg, Huh:2023jxt}. In fact, it was shown that
\begin{equation}
\ddot{\cal C}_K(t)= 2\sum_{n\geq 0}(b_{n+1}^2- b_n^2)|\varphi_n(t)|^2 + (a_{n+1}- a_n)b_{n+1} \left(\varphi_n^* \varphi_{n+1}+ \varphi_{n+1}^* \varphi_n \right).  \label{pedrogonzalez} 
\end{equation}
In this expressions $a_n, b_n$ are the Lanczos coefficients and $\varphi_n$ the coefficients of the state $|\psi(t)>$ in an expansion in terms of the Krylov basis $|K_n>$, see around eq.(\ref{betoalonso}). In some cases, it occurs that the coefficients $a_n$ are constant, independent of $n$. In such cases eq.(\ref{pedrogonzalez}) tells that the sequence of $b_n$ has the same behaviour as the acceleration of the complexity. In the cases analysed in this work, we have a 'convex' sequence, for which $b_{n+1}>b_n$.

More on Lanczos coefficients. At short times, the Krylov spread complexity has an expansion in even powers \cite{Muck:2026top, Balasubramanian:2022tpr, Caputa:2021sib}. Namely, for short times
\begin{equation}
{\cal C}_K(t)=    b_1^2 t^2+ \frac{b_1^2}{12}\left[ 2 b_2^2-4b_1^2-(a_1-a_0)^2\right]t^4 + {\cal O}(t^6).
\end{equation}
We can use this (and higher orders in the expansion that are easy to obtain) to bridge the geometry results in this paper with the Lanczos coefficient $b_n, a_n$. In fact,
for $t=0$ we find
\begin{equation}
b_1^2= \frac{m\sqrt{A(r_{UV})}}{4} {\cal U}(r_{UV}).    
\end{equation}
In this expression it becomes useful the expression for ${\cal U}$ in eq.(\ref{udet}). Similarly, using the expression for $\dot{r}$ in eq.(\ref{eq:rdot_Krylov}) and the chain rule, we find
\begin{eqnarray}
 & & \frac{\ddot{\cal U}(0)}{{\cal U}(0)}= 2 b_2^2-4b_1^2-(a_1-a_0)^2=  \frac{\ddot{\cal U}(r_{UV})}{{\cal U}(r_{UV})}=-\frac{A'(r_{UV}) \frac{d~{\cal U}(r_{UV})}{d~r}}{2 A(r_{UV}) B(r_{UV}) {\cal U}(r_{UV})} .\label{patofillol}
\end{eqnarray}
In the forthcoming work, we apply these expressions and others to start mapping the Lanczos coefficients using holography. This also serves to propose a survival amplitude satisfying the constraints required by the symmetries of the QFT. We leave this  to be studied in the future.

\bibliographystyle{JHEP}
\bibliography{main.bib}

\end{document}